\documentclass[12pt]{article}
\usepackage[authoryear]{natbib}
\usepackage{graphicx}
\usepackage{amssymb,amsthm,amsmath,epsfig}
\usepackage{times}
\usepackage{bm}
\usepackage{color}
\usepackage{setspace}
\usepackage[utf8]{inputenc}
\usepackage{lscape}
\usepackage[nottoc]{tocbibind}
\usepackage[toc,page]{appendix}

\newcommand\x{2.5}
\usepackage[top = \x cm, bottom = \x cm, left = \x cm,right = \x cm]{geometry}

\usepackage{authblk}

\title{Mortality modeling at old-age: a mixture model approach}

\author[1]{Silvio C. Patricio\thanks{silca@sam.sdu.dk}}
\author[2]{Fredy Castellares}
\author[3]{Bernardo L. Queiroz}
\affil[1]{\small{The Interdisciplinary Centre on Population Dynamics, University of Southern Denmark}}
\affil[2]{\small{Departamento de Estatística, Universidade Federal de Minas Gerais}}
\affil[3]{\small{Departamento de Demografia, Universidade Federal de Minas Gerais}}

\begin{document}
\date{}
\maketitle

\begin{abstract}
% \noindent In this paper, we propose a mixture-based model for mortality modeling above age 70. The proposed model is compared with 4 other widely used models: the Beard, Gompertz, Makeham, and Perks models. Our model captures well the mortality rate's behavior at all the ages. We applied the method to a country with high quality data, Japan, and one with lower data quality, Brazil. In the comparative study for the Japanese population, the model presented a better fit to the data, obtaining an absolute mean percentage error of less than 7\%, while the other models presented values greater than 30\%. \\

\subsubsection*{BACKGROUND}
Estimates of mortality at older ages, even above age 70, are a major concern for demographers and have important applications and consequences in other areas such as actuarial science and economics. In general, mortality estimates at older ages are limited by small numbers both in the exposure and events 
\subsubsection*{OBJECTIVE}
In this paper, we propose a mixture-based model for mortality modeling for the elderly (+70 years).
\subsubsection*{METHODS}
The proposed model is compared with 4 other widely studied and used models: the Beard, Gompertz, Makeham, and Perks models. We apply our proposed method to two populations of different data quality: Brazil and Japan.
\subsubsection*{RESULTS}
The mixture-based model captures the decrease in mortality force at older ages, which is a characteristic observed in several populations.
\subsubsection*{CONCLUSIONS}
In the comparative study for the Japanese population, our model presented a better fit to the data, obtaining an absolute mean percentage error of less than 7\%, while the other models presented values greater than 30\%.
\subsubsection*{CONTRIBUTION}
In the comparative study for the Japanese population, our model presented a better fit to the data, obtaining an absolute mean percentage error of less than 7\%, while the other models presented values greater than 30\%.
\\
\\
\noindent{\em Keywords}: old-age; mortality modeling; mixture model; Beard model; Gompertz model; Gompertz-Makeham model; Perks model.
\end{abstract}
\newpage

\section{Introduction}
In the past centuries, much has been done to model the process of mortality in populations and its consequences \citep{graunt1977natural, gompertz1825xxiv, wilmoth2000demography, van2021have}. One of humanity’s most outstanding achievements in the last century, perhaps the last millennium, has been the four-decade increase in human life expectancy over the past 160 years \citep{vaupel2021demographic, wilmoth2000demography} and the improvement in human mortality. All these changes in human longevity directly affect pension, welfare, and health care systems \citep{cutler2006determinants}. 

Despite pioneering work by \citeauthor{graunt1977natural} and \citeauthor{gompertz1825xxiv}, understanding of mortality for older ages remains a challenge, specially in developing countries with more defective data. In general, mortality estimates at older ages are limited by small numbers both in the exposure, death count and problems with age declaration \citep{feehan2018separating, wrigley2014mortality, nepomuceno2019population}. There is an important and ongoing debate about the levels of mortality at older ages. In general terms, the debate is whether mortality at older ages is declining or continues to increase \citep{gavrilov2019late, feehan2018separating}. In some settings, such as Brazil, there is also an important question on the crossover of mortality at older ages when comparing different population sub-groups \citep{nepomuceno2019population, pinheiro2019regional, gomes2009number}.

In addition to the problem of the quality of the data, there is a debate on hypotheses of selectivity and of the biological limit of mortality in human populations that, in different ways, would impact the behavior of mortality taxes in more advanced ages. One of the consequences of the mortality selectivity hypothesis would be a greater rate of deceleration of the rates of mortality in more advanced ages. In this context, there are a series of models to explain mortality behavior at older ages. The choice of the appropriate model depends on the hypotheses assumed, whether in relation to the quality of the two data or in relation to the impacts produced by the selectivity.

There are several possible explanations for the observed results and estimates. First one is related to data quality in different areas of a country, across sub-population groups and age. For instance, it could be a consequence of different age misreporting patterns or issues with quality of vital registration systems \citep{black2017methuselah}. Preston et al (2000) investigated how different types of age misreporting can affect estimates of mortality rates at older ages, by analyzing the effects of three patterns of age misreporting: net age overstatement, net age understatement, and symmetric age misreporting.. It is also possible that mortality selection plays a role in the observed levels of mortality at older ages \citep{barbi2018plateau, wachter2018hypothetical}. In the context of higher mortality rates at young ages, survivors to older ages would be physiologically stronger and then live longer than others. 

Unfortunately, data quality at older ages limits the understanding of mortality and the evolution of survivorship at older ages. \cite{feehan2018separating} uses alternative methods to cohort mortality above age 80. He finds that no model can be universally applied to estimate old-age mortality, but he argues that Log-Quad \citep{wilmoth2012flexible} provides a good fit. However, the log-quad method is based on standard mortality changes from the Human Mortality Database that is constructed from a series of countries in the Northern Hemisphere and might be limited to low and middle income countries. 

In this paper, we suggest a model that captures decline in mortality rates at older ages, which is a characteristic observed in some populations. Based on the proposed model, we perform a comparative study using establish mortality laws with our proposed approach.  
The analysis was split into two parts. First, to compare the four widely used models with the proposed model: in this part we will study the behavior of these models in two databases: one with good quality data on mortality in Japan in 2015 (obtained from The Human Mortality Database of mortality), and the other database that has limited data regarding mortality in Brazil in 2010. In it the models will be evaluated from Mean Absolute Percentage Error (MAPE) of the log-hazard using the leave-one-out cross-validation method, and the model with the least MAPE will all be the best model. Moreover, as some models are complex, the genetic algorithm was used to obtain the estimates via maximum likelihood. Using this algorithm ensures convergence to the global maximum value. The second part applies the proposed model to different databases, and aims to understand the model behavior and also to verify its potential for application to real data.The model presented a better fit to the data, obtaining an absolute mean percentage error of less than 7\%, while the other models presented values greater than 30\%.

\section{Models specification's and parameter estimation}
Considering a non negative random variable (r.v.) $T$ defined in a probability space $(\mathbb{R}_+, \mathcal{B}, \mathbb{P}_{\bm{\theta}})$, representing the individual life-spam, the r.v. $T$ can be characterized by the survival function
\begin{equation*}
    S(x|{\bm{\theta}}) = \mathbb{P}_{\bm{\theta}}(T>x)
\end{equation*}{}
\noindent which is associated with the density
\begin{equation*}
    f(x|{\bm{\theta}}) = -\frac{\partial}{\partial x} S(x|{\bm{\theta}}).
\end{equation*}{}
If $ S $ is a continuous survival function associated with a $ f $ density function, then the function $\mu$ defined in $\mathbb{R}_+$ by
\begin{equation*}
    \mu(x|{\bm{\theta}}) = \lim_{\varepsilon \downarrow 0} \frac{\mathbb{P}_{\bm{\theta}}(x<T<x+\varepsilon | X>x)}{\varepsilon} =  \frac{f(x|{\bm{\theta}})}{S(x|{\bm{\theta}})}
\end{equation*}{}
\noindent it's called the $ T $ mortality force. This function is usually used to describe the force of mortality for a group of people or population.

The inferences in the model are based on the assumption that the number of death has a Poisson distribution. Therefore, be $\mathbf{D}=(D_0, D_1, \dots, D_m)'$ a random sample with Poisson distribution, with $D_k$ representing the number of deaths between ages $[k, k+1)$, with $k=0, \dots, m$, i.e. the number of death of people with $ k $ years old. 

For this approach it is considered that $\mathbb{E}(D_k)=\mu(k| \bm{\theta}) E_k$, with $\mu(k|  \bm{\theta})$ representing the mortality force at age $k$, where $\bm{\theta}=(\theta_1, \theta_2, \dots, \theta_p)'$ is the parameter vector that characterizes the mortality rate, and $E_k$ the population at age $k$ exposed to risk, that are assumptions widely used by demographers \citep{brillinger1986natural}. Also, as it is the Poisson distribution, we have to $\mathbb{V}(D_k)=\mu(k|  \bm{\theta}) E_k$, same value of expectation.

Be $\bm{D}=(D_0, \dots, D_m)'$ e $\bm{E}=(E_0, \dots, E_m)'$. The log-likelihood function from $\bm{\theta}$ is given by
\begin{equation}
    \ell (\bm{\theta} | \mathbf{D} ) = \sum_{k=1}^m D_k\log\lambda(\bm{\theta}, k) -\lambda(\bm{\theta}, k), \label{eq:log_vero}
\end{equation}{}

\noindent with $\lambda(\bm{\theta}, x) = \mu(x|  \bm{\theta}) E(x)$. The likelihood estimate $\widehat{\bm{\theta}}$ is obtained from maximizing the log-likelihood function with in equation \ref{eq:log_vero}, with respect to $\bm{\theta}$. Obtaining the partial derivative vector of the equation \ref{eq:log_vero}, with respect to $\theta_i$, $i=1, \dots, p$, we have
\begin{equation}
    \frac{\partial\ell (\bm{\theta} | \mathbf{D} )}{\partial{\theta_i}} = \sum_{k=1}^m
    \left( \frac{D_k}{\mu(k|  \bm{\theta})} -E_k\right)\frac{\partial \mu(k|  \bm{\theta})}{\partial{\theta_i}}. \label{eq:grad}
\end{equation}{}
\noindent The likelihood estimation can also be obtained by equating the partial derivative vector to zero and simultaneously solving the system of equations. The explicit form of the gradient vector is explained for each of the models considered in this article. The Newton-Raphson method can be applied to solve the likelihood equation to obtain the estimate $\widehat{\bm{\theta}}$.

% -----------------------------------------
% modelo de beard
% -----------------------------------------
\subsection{Beard model} % R. E Beard (1959)
In this model introduced in \cite{beard1959note}, we have that the force of mortality is given by
\begin{equation*}
    \mu(k|  \bm{\theta}) = \frac{ae^{bk}}{1+\delta e^{bk}}
\end{equation*}{}
\noindent with $\bm{\theta} = (a, b, \delta)'\in \mathbb{R}_+^3$. From which we calculate the partial derivative with respect to $ a $ and $ b $. E Equation \ref{eq:grad} gives us a general equation for the gradient vector, where it depends only on the mortality rate and its partial derivative with respect to each parameter. Hence we get
\begin{align*}
    \frac{\partial\ell (\bm{\theta} | \mathbf{D} )}{\partial{a}} &= \sum_{k=1}^m
    \left[ D_k\left(\frac{1+\delta e^{bk}}{ae^{bk}}\right) -E_k\right]\frac{e^{bk}}{\left( 1+\delta e^{bk} \right)}\\
    \frac{\partial\ell (\bm{\theta} | \mathbf{D} )}{\partial{b}} &= \sum_{k=1}^m
    \left[ D_k\left(\frac{1+\delta e^{bk}}{ae^{bk}}\right) -E_k\right]\frac{ake^{bk}}{\left( 1+\delta e^{bk} \right)^2}\\
    \frac{\partial\ell (\bm{\theta} | \mathbf{D} )}{\partial{\delta}} &= \sum_{k=1}^m
    \left[ D_k\left(\frac{1+\delta e^{bk}}{ae^{bk}}\right) -E_k\right]\frac{ae^{2bk}}{\left( 1+\delta e^{bk} \right)^2}
\end{align*}{}
\noindent representing the gradient vector.
% -----------------------------------------
% modelo de Gompertz
% -----------------------------------------
\subsection{Gompertz model}
In this model introduced in \cite{gompertz1825}, we have that the force of mortality is given by
\begin{equation*}
    \mu(k| \bm{\theta}) = ae^{bk},
\end{equation*}{}
\noindent with $\bm{\theta} = (a, b)' \in \mathbb{R}_+^2$. So for the gradient vector we have
\begin{align*}
    \frac{\partial\ell (\bm{\theta} | \mathbf{D} )}{\partial{a}} &= \sum_{k=1}^m
    \left[ \frac{D_k}{ae^{bk}} -E_k\right]e^{bk}\\
    \frac{\partial\ell (\bm{\theta} | \mathbf{D} )}{\partial{b}} &= \sum_{k=1}^m
    \left[ \frac{D_k}{ae^{bk}} -E_k\right]ake^{bk}
\end{align*}{}
% -----------------------------------------
% modelo Makeham
% -----------------------------------------
\subsection{Makeham model}
In this model introduced in \cite{makeham1860law}, we have that the force of mortality is given by
\begin{equation*}
    \mu(k| \bm{\theta}) = ae^{bk}+c,
\end{equation*}{}
\noindent with $\bm{\theta} = (a, b, c)' \in \mathbb{R}_+^3$. So for the gradient vector we have
\begin{align*}
    \frac{\partial\ell (\bm{\theta} | \mathbf{D} )}{\partial{a}} &= \sum_{k=1}^m
    \left[ \frac{D_k}{ae^{bk}+c} -E_k\right]e^{bk}\\
    \frac{\partial\ell (\bm{\theta} | \mathbf{D} )}{\partial{b}} &= \sum_{k=1}^m
    \left[ \frac{D_k}{ae^{bk}+c} -E_k\right]ake^{bk}\\
    \frac{\partial\ell (\bm{\theta} | \mathbf{D} )}{\partial{c}} &= \sum_{k=1}^m
    \left[ \frac{D_k}{ae^{bk}+c} -E_k\right]
\end{align*}{}
% -----------------------------------------
% modelo Perks
% -----------------------------------------
\subsection{Perks model} % Perks (1932)
In this model introduced in \cite{perks1932some}, we have that the force of mortality is given by
\begin{equation*}
    \mu(k| \bm{\theta}) = \frac{\gamma + ae^{bk}}{1+\delta e^{bk}}
\end{equation*}{}
\noindent with $\bm{\theta} = (a, b, \gamma, \delta)'$. So for the gradient vector we have
\begin{align*}
    \frac{\partial\ell (\bm{\theta} | \mathbf{D} )}{\partial{a}} &= \sum_{k=1}^m
    \left[ D_k\left(\frac{1+\delta e^{bk}}{\gamma+ae^{bk}}\right) -E_k\right]\frac{e^{bk}}{1+\delta e^{bk} }\\
\end{align*}
\begin{align*}
    \frac{\partial\ell (\bm{\theta} | \mathbf{D} )}{\partial{b}} &= \sum_{k=1}^m
    \left[ D_k\left(\frac{1+\delta e^{bk}}{\gamma+ae^{bk}}\right) -E_k\right]\frac{k(a-\delta\gamma)e^{bk}}{\left(1+\delta e^{bk}\right)^2 }\\
    \frac{\partial\ell (\bm{\theta} | \mathbf{D} )}{\partial{\gamma}} &= \sum_{k=1}^m
    \left[ D_k\left(\frac{1+\delta e^{bk}}{\gamma+ae^{bk}}\right) -E_k\right]\frac{1}{1+\delta e^{bk}}\\
    \frac{\partial\ell (\bm{\theta} | \mathbf{D} )}{\partial{\delta}} &= \sum_{k=1}^m
    \left[ D_k\left(\frac{1+\delta e^{bk}}{\gamma+ae^{bk}}\right) -E_k\right]\frac{e^{bk}\left(ae^{bk}+\gamma \right)}{\left(1+\delta e^{bk}\right)^2 }
\end{align*}{}

% -----------------------------------------
% modelo Weibull 
% -----------------------------------------
%\subsection{Weibull} % Weibull (1951)
%Neste modelo introduzido em \cite{weibull1951statistical}, temos que a %taxa de mortalidade é dada por
%\begin{equation*}
%    \mu(k| \bm{\theta}) = ak^{b-1}
%\end{equation*}{}
%\noindent onde $\bm{\theta} = (a, b)'$. Portanto, para o vetor gradiente %obtemos
%\begin{align*}
%    \frac{\partial\ell (\bm{\theta} | \mathbf{D} )}{\partial{a}} &= %\sum_{k=1}^m
%    \left[ \frac{D_k}{ae^{bk}} -E_k\right]k^{b-1}\\
%    \frac{\partial\ell (\bm{\theta} | \mathbf{D} )}{\partial{b}} &= %\sum_{k=1}^m
%    \left[ \frac{D_k}{ae^{bk}} -E_k\right]ak^{b-1}\log k
%\end{align*}{}

% -----------------------------------------
% modelo de mistura 
% -----------------------------------------
\subsection{Mixture model}

As with \citeauthor{makeham1860law}, we will seek to decompose mortality into two components: premature and senescent mortality, respectively modeled by an exponential and a Gompertz component. However, \citeauthor{makeham1860law} distinguishes these components through mortality force, and here we propose to distinguish them through distribution. Therefore, we are considering that the r.v. $ T $  introduced at the beginning of this session is associated with a probability density function $ f $, which is define as:
\begin{equation}
    f(x|\bm{\theta}) = p\left( \lambda e^{-\lambda x} \right)+(1-p) \left( ab\exp\left\{\-a\left(e^{bx}-1 \right)+bx \right\} \right) \label{eq:dist_mix}
\end{equation}{}
\noindent with $\bm{\theta} = (a, b, \lambda, p)'$.

The density $ f $ is a Gompertz and a exponential distribution a mixture. The Gompertz distribution will fit the senescence deaths count, and the exponential distribution will fit the premature deaths, such as accidents and disease. Briefly, this model considers the existence of two sub populations in the death count, one Gompertz and the other Exponential, and the parameters $ p $ and $ q = 1-p $ represent the proportions of each one.

Since the random variable $ T $ is associated with a density function, we can also associate it with a hazard function. In this case the force of mortality is defined by:
\begin{equation}
    \mu(x|\bm{\theta}) = \frac{f(x|\bm{\theta})}{S(x|\bm{\theta})}=\frac{p\left( \lambda e^{-\lambda x} \right)+(1-p) \left( ab\exp\left\{\-a\left(e^{bx}-1 \right)+bx \right\} \right)}{pe^{-\lambda x}+(1-p)\exp\{-a\left( e^{bx}-1\right)\}}, \label{eq:taxa_mix}
\end{equation}{}
\noindent for which there is no straightforward interpretation. Which is lost due to the ease of deriving functions such as statistical moments and expected average residual life (for more details, see \cite{finkelstein2009understanding}) .From this we can get the gradient vector which, for this model, is given by
\begin{align*}
    \frac{\partial\ell (\bm{\theta} | \mathbf{D} )}{\partial{a}} &= \sum_{k=1}^m
    \left[ D_k\frac{pe^{-\lambda k}+(1-p)\exp\{-a\left( e^{bk}-1\right)\}}{p\left( \lambda e^{-\lambda k} \right)+(1-p) \left( ab\exp\left\{\-a\left(e^{bk}-1 \right)+bk \right\} \right)} -E_k\right] \times\\
    &\times \frac{b(1-p)e^{a(ke^b-1)+b\lambda}+ab(1-p)(ke^b-1)e^{a(ke^b-1)+bx}}{pe^{-\lambda k}+(1-p)\exp\{-a\left( e^{bk}-1\right)\}}+ \\
    &+  (-1)  \frac{(1-p)(1-e^{bk})e^{-a(e^{bk}-1)}\left( p\left( \lambda e^{-\lambda k} \right)+(1-p) \left( ab\exp\left\{\-a\left(e^{bk}-1 \right)+bk \right\} \right)\right)}{\left( pe^{-\lambda k}+(1-p)\exp\{-a\left( e^{bk}-1\right)\} \right)^2} \\
    \frac{\partial\ell (\bm{\theta} | \mathbf{D} )}{\partial{b}} &= \sum_{k=1}^m
    \left[ D_k\frac{pe^{-\lambda k}+(1-p)\exp\{-a\left( e^{bk}-1\right)\}}{p\left( \lambda e^{-\lambda k} \right)+(1-p) \left( ab\exp\left\{\-a\left(e^{bk}-1 \right)+bk \right\} \right)} -E_k\right] \times\\
    &\times \frac{a(1-p)xe^{bx-a(e^{bx}-1)}\left( ab(1-p)e^{a(xe^{bx}-1)+bx}+\lambda p e^{-\lambda x}\right)}{\left((1-p)e^{a(e^{bx}-1)}+pe^{-\lambda x} \right)^2}+\\
    &  +\frac{a(1-p)e^{a(e^{bx}-1)+bx}+ab(1-p)e^{a(e^{bx}-1)+bx}\left( axe^b + x \right)}{(1-p)e^{a(e^{bx}-1)}+pe^{-\lambda x}} \\
    \frac{\partial\ell (\bm{\theta} | \mathbf{D} )}{\partial{\lambda}} &= \sum_{k=1}^m
    \left[ D_k\frac{pe^{-\lambda k}+(1-p)\exp\{-a\left( e^{bk}-1\right)\}}{p\left( \lambda e^{-\lambda k} \right)+(1-p) \left( ab\exp\left\{\-a\left(e^{bk}-1 \right)+bk \right\} \right)} -E_k\right] \times\\
    &\times \left[\frac{p e^{-\lambda x}-\lambda p x e^{-\lambda x}}{(1-p) e^{-a \left(e^{b x}-1\right)}+p e^{-\lambda
   x}}+\frac{p x e^{-\lambda x} \left(a b (1-p) e^{a \left(x e^b-1\right)+b x}+\lambda p
   e^{-\lambda x}\right)}{\left((1-p) e^{-a \left(e^{b x}-1\right)}+p e^{-\lambda x}\right)^2} \right]\\
   \frac{\partial\ell (\bm{\theta} | \mathbf{D} )}{\partial{p}} &= \sum_{k=1}^m
    \left[ D_k\frac{pe^{-\lambda k}+(1-p)\exp\{-a\left( e^{bk}-1\right)\}}{p\left( \lambda e^{-\lambda k} \right)+(1-p) \left( ab\exp\left\{\-a\left(e^{bk}-1 \right)+bk \right\} \right)} -E_k\right] \times\\
    &\times \left[\frac{\lambda e^{-\lambda x}-a b e^{a \left(x e^b-1\right)+b x}}{(1-p) e^{-a \left(e^{b
   x}-1\right)}+p e^{-\lambda x}}-\frac{\left(e^{-\lambda x}-e^{-a \left(e^{b x}-1\right)}\right)
   \left(a b (1-p) e^{a \left(x e^b-1\right)+b x}+\lambda p e^{-\lambda x}\right)}{\left((1-p)
   e^{-a \left(e^{b x}-1\right)}+p e^{-\lambda x}\right)^2}\right]
\end{align*}{}

\section{Data and empirical results}

In order to evaluate the proposed model, we will compare its performance on high and low-quality data. For this, we will evaluate its performance against four other models, using the Mean Absolute Percentage Error (MAPE) combined with the leave-one-out cross-validation method, which will measure the average distance between the log-hazard and the log-mortality rate.  Moreover, as some models are highly nonlinear, the Genetic Algorithm \citep{alg_genetic, mirjalili2019genetic} will be used to maximize the likelihood function. This algorithm ensures convergence to the global maximum value.

\subsection{Models comparison}
\subsubsection*{In a high quality data setting}

\begin{figure}[htb!]
    \centering
    \includegraphics[height = 6cm]{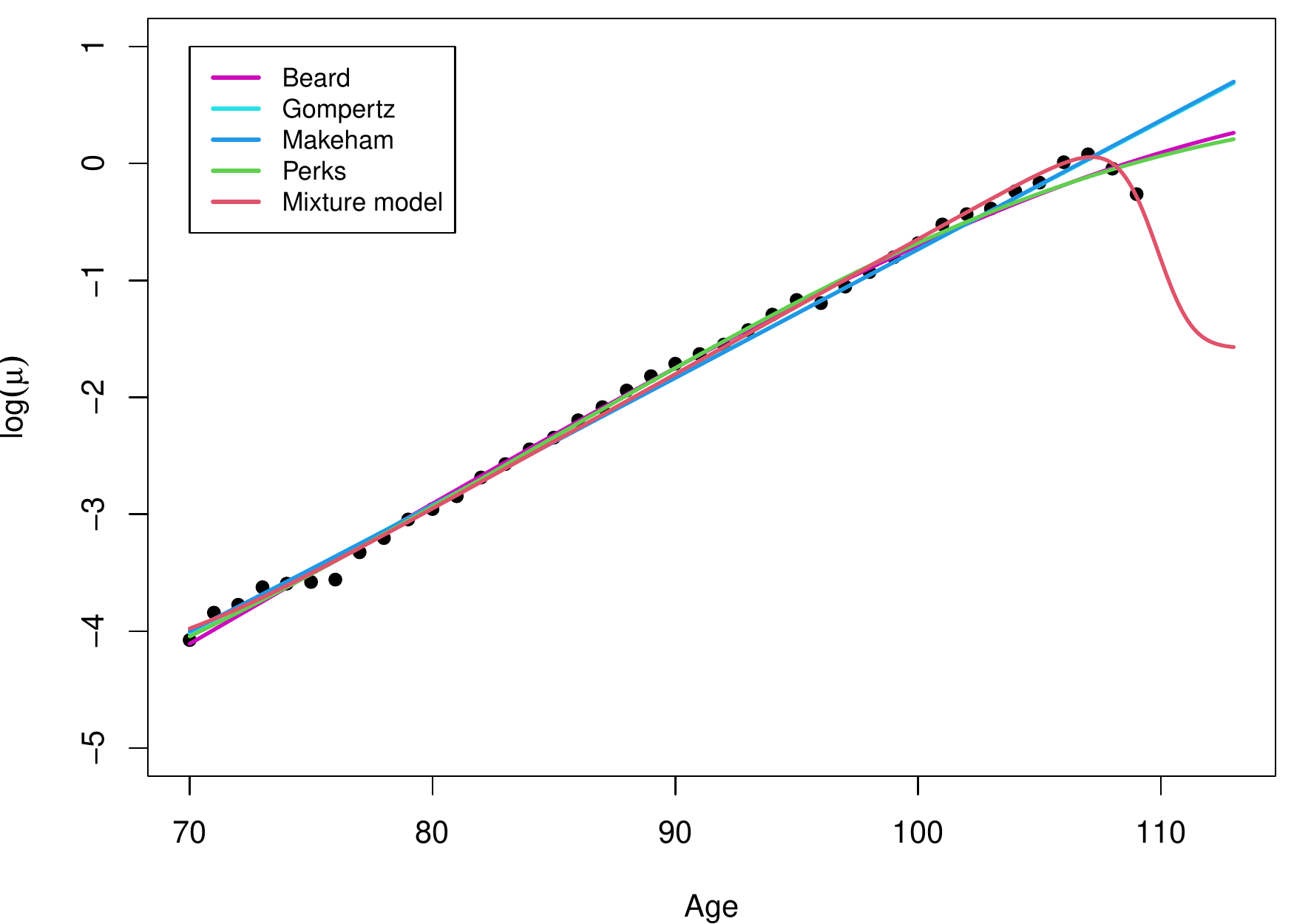}
    \caption{Japan 2015 modeling}
    \label{fig:comparison}
\end{figure}

In this scenario, we will use mortality data from Japan in 2015 obtained from The Human Mortality Database (HMD). The observed value of $\log \mu$ is linearly increasing to a certain age, and then has a sharp drop. This behavior was also noted this country in the last three decades. However this is not restricted to Japan, other countries like Sweden, Germany, USA and Korea also had the same mortality behavior. The Figure \ref{fig:comparison} shows the estimated log-hazard function. We can clearly see the models of Beard, Gompertz, Makeham and Perks were not able to fit properly the mortality rate after age 100.

{}

The Gompertz model consider force of mortality being log-linear, but clearly this behavior does not describe the entire observed curve. For this model the estimated parameter is $\widehat{\bm{\theta}} = (0.0179, 0.1094)'$. And this model has a MAPE of 34.0127, i.e., this model's predictions are on average 34.0127 \% distant of the observed value. A similar result can be obtained from the Makeham model, which has estimated parameter $\widehat{\bm{\theta}} = (0.0174 ,0.1103 ,0.0008)'$, and MAPE 33.0288.

The Beard can be seen as the ratio of a Gompertz and a Makeham models with $c = 1$, with the parameters estimated by ML $\widehat{\bm{\theta}} = (0.0165, 0.1216, 0.0073)'$. Despite Beard's combination of Makeham and Gompertz models, this model provided the worse fit, reaching a MAPE of 55.6189.

The Perks model also has a similar construction to Beard. It is the ratio between two Makeham models. For this model we estimate $\widehat{\bm{\theta}} = (0.0135, 0.1313 ,0.0040 ,0.0075)'$. And as expected, this model had a very similar behavior to the previous model, including in MAPE of 51.3591, suggesting that this model does not fit well to the data.

Finally, for the proposed mixture-based model, we estimated $\widehat{\bm{\theta}} = (0.1155, 0.0163, 0.2061, 0.0126)'$, and a MAPE of 6.9193, the best of the models presented in this study. In addition, this model was the only one that was able to capture the sharp drop in the mortality rate. With the estimated parameters we can interpret that the non-senescence death represents 1.2599 \% of the total death after age 70.

%\subsection{Models comparison in limited data}
\subsubsection*{In a low quality data setting}

We observed that the model works well on data that has good quality, and now we aim to understand how the model behaves when the data has limitations. In this case we are going to use data from Brazil from 2010 \citep{queiroz2020comparative, gonzaga2016estimating} . Previous studies showa a mortality crossover above age 60 when comparing more and less developed states in Brazil using the Topals model \citep{queiroz2020comparative, gonzaga2016estimating}. It is argued that the result is related to the level of completeness of death counts, age misreporting and mortality selection. Thus, it is an important and relevant case study for the application of our prosed mixture model. For this, as before, we will compare the performance of the 5 models presented through MAPE.\\

\begin{figure}[htb!]
    \centering
    \includegraphics[height = 6cm]{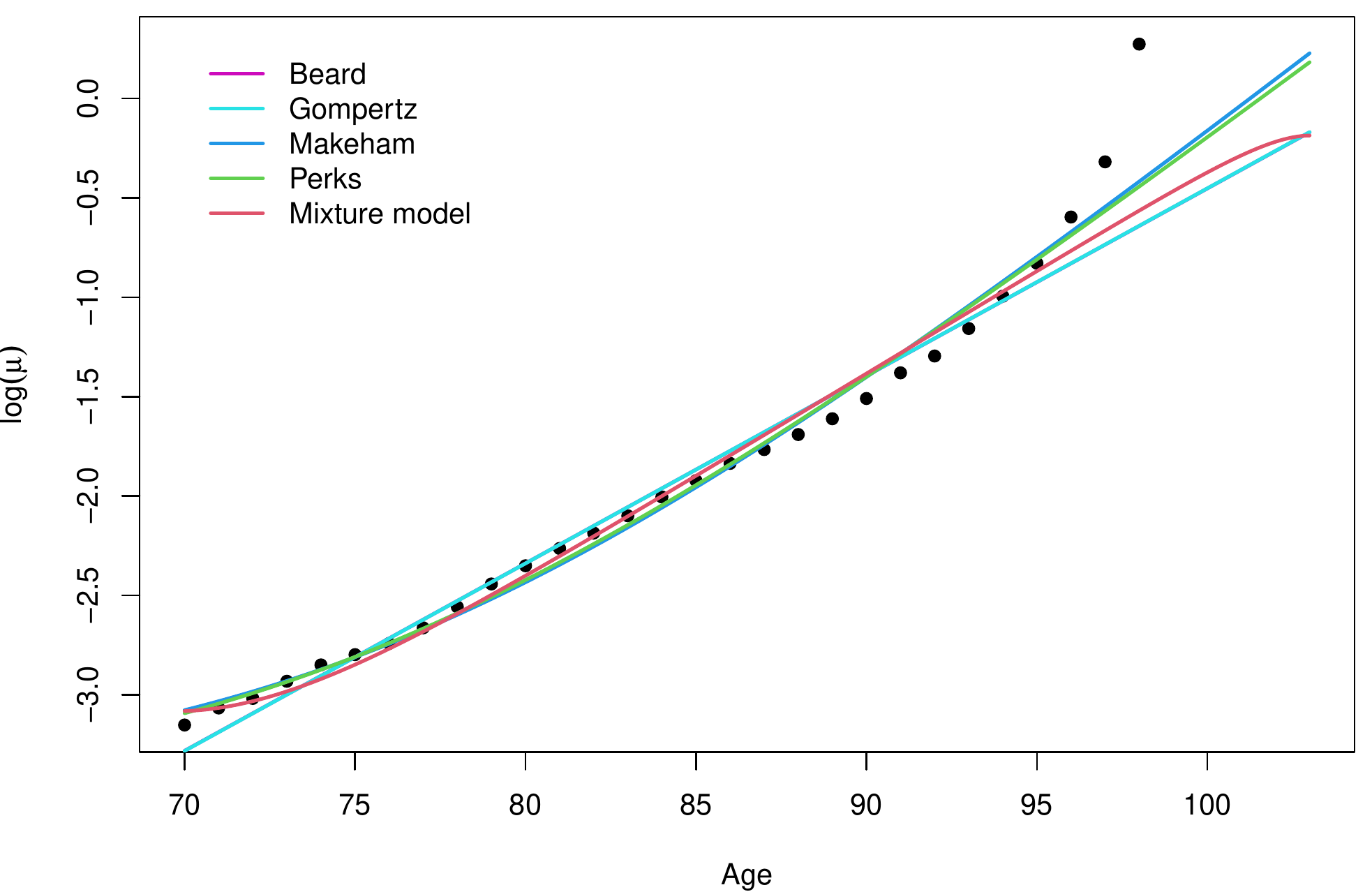}
    \caption{Brazil 2010 modeling}
    \label{fig:application_brazil}
\end{figure}{}

For the first model (Beard) we estimated $\widehat{\bm{\theta}} = \left(0.0375, 0.0942, 5.5625 \times 10^{-8}  \right)'$, and a MAP of 20.4629, i.e., on average this model distanced by 20 \% of the mortality rate. We also got a similar conclusion about the Gompertz model, estimating $ \widehat{\bm{\theta}} = (0.0375, 0.0943)'$ and MAPE about 20.4499.

The Makeham and Perks models also obtained similar results. For Makeham it was estimated $\widehat{\bm{\theta}} = \left(0.01481, 0.1338, 0.03131\right)'$ resulting in a MAP of 14.5473, and for Perks model it was estimated $\widehat{\bm{\theta}} = \left(0.0163, 0.0129, 0.0290, 3.4272 \times 10^{-7}  \right)'$ which results in MAPE of 14.9002.

Finally, for the proposed model was estimated $\widehat{\bm{\theta}} = (0.1036, 0.0315, 0.2389, 0.0692)'$, and a MAPE of 18.0038\%, which indicates that the model is not able to capture mortality well in these data. Therefore, the results found in this application match the results discussed in \cite{feehan2018separating} on the power of models capturing mortality at advanced ages universally.
 
\subsection{Model applications}
\begin{figure}[htb!]
    \centering
    \includegraphics[width = 17cm]{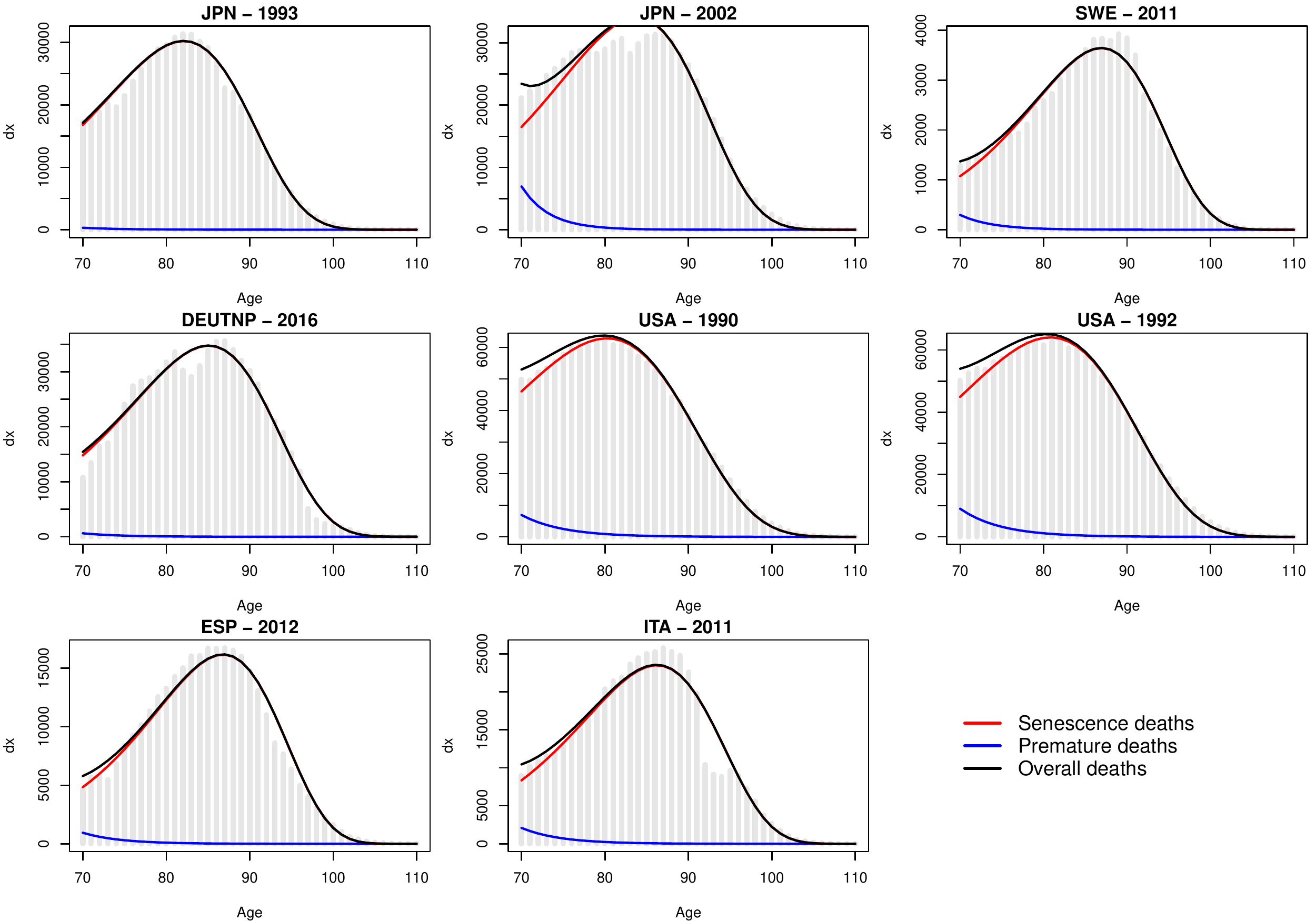}
    \caption{Estimations}
    \label{fig:application2}
\end{figure}

As we have seen, the proposed model has a high capacity to fit the mortality at older ages. Therefore, we will illustrate the power of this model by applying it to mortality data from Japan (1993 and 2002), Sweden (2011), Germany (2016), USA (1990 and 1992), Spain (2012) and Italy (2011). Table \ref{tab:estim} represents the estimate for each dataset, and Figure \ref{fig:application2} represents their respective decomposed distribution of death.

\begin{table}[!htp]
\centering
{\small
\caption{Parameters estimated.}\label{tab:estim}
\begin{tabular}{ccccccr}
\hline
Country & Year & $\hat a$ & $\hat b$ & $\hat c$ & $\hat p$ & MAPE\\ 
  \hline
Japan & 1993 & 0.10911 & 0.02916 & 0.21615 & 0.00250 & 8.86459\\ 
Japan & 2002 & 0.10897 & 0.02425 & 0.30152 & 0.03276 & 7.49451\\ 
Sweden & 2011 & 0.12390 & 0.01520 & 0.26448 & 0.01559 & 12.27019\\ 
Germany & 2016 & 0.11046 & 0.02090 & 0.22283 & 0.00397 & 10.68258\\ 
USA & 1990 & 0.08845 & 0.03569 & 0.20360 & 0.02569 & 3.80694\\ 
USA & 1992 & 0.09057 & 0.03404 & 0.20575 & 0.03217 & 2.91887\\ 
Spain & 2012 & 0.12372 & 0.01544 & 0.22751 & 0.01307 & 12.38755\\ 
Italy & 2011 & 0.11606 & 0.01768 & 0.21710 & 0.01999 & 13.24385\\ 
\hline
\end{tabular}    }
\end{table}

In Table \ref{tab:estim} it can be seen that the estimated values for $p$ are small, less than 0.04, which indicates that the proportion of premature deaths above age 70 does not exceed 4\%. This result was already expected, since by truncating the mortality data at age 70, we are excluding infant mortality and mortality hump \citep{remund2018cause}, and we only observe the tail of the distribution of premature mortality. Furthermore, the our result is also in agreement with \citeauthor{horiuchi1997age}'s results, that above age 75 mortality decelerates for most causes of death \citep{horiuchi1997age}.

\begin{figure}[htb!]
    \centering
    \includegraphics[width = 10cm]{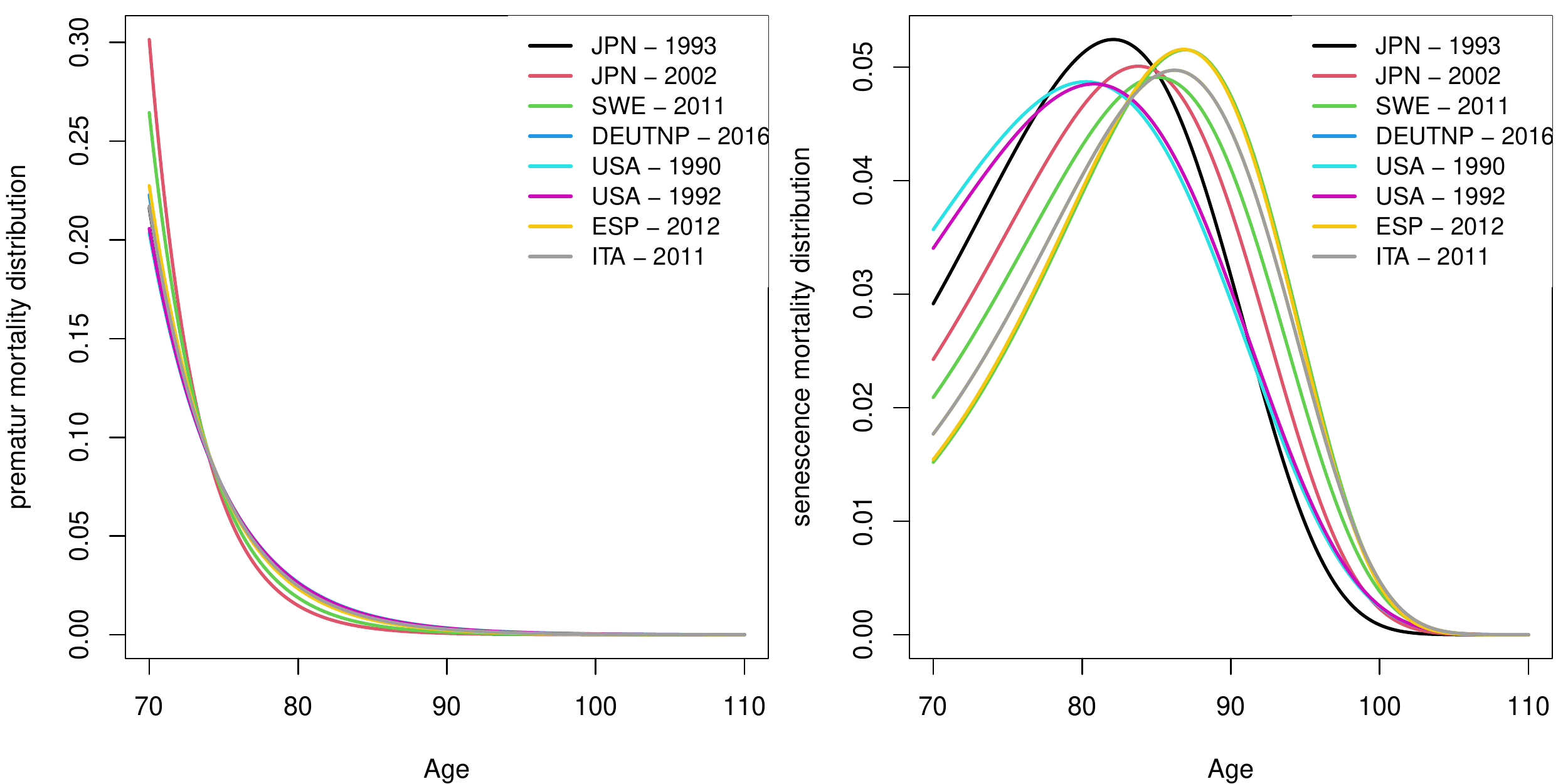}
    \caption{Estimates of mortality components.}
    \label{fig:application1}
\end{figure}{}

The estimated values for the $c$ parameter are similar, and concentrated around 0.23. This suggests that, despite having different proportions, the distributions of premature death are similar, as can be seen on the left in Figure \ref{fig:application1}. Such similarity was not observed in the senescent death distributions, which have a marked difference, as can be seen on the right in Figure \ref{fig:application1}. Despite this, it is clear that the modal age of death is between 80 and 90, which is consistent with previous studies and \citep{horiuchi2013modal}.

{}

The Figure \ref{fig:application2} shows the distribution of death estimated and broken down into premature and senescent deaths. In it, we can observe the quality of fit of the estimated model (black line). In addition, it is possible to see that for Japan in 1993 and Germany in 2016, there were practically no premature deaths after age 70, this could also be inferred from analyzing the Table \ref{tab:estim}, where the values of estimate for $p$ are small.

\section{Conclusions and future works}

Robust estimates of mortality rates in advanced ages are a challenge for demographers for various reasons. Even in populations with good records of deaths and population there are disturbances in the function of the low number of events and/or some limitation in the information on the age of death. In case of countries where the problems of data quality is present, the challenges are greater. 

For some centuries there has been an ambition to decompose mortality into interpretable components. The best known are those proposed by \cite{makeham1860law} and \cite{pollard_model}. However, in recent years researchers have devoted to this problem \citep{remund2017analyzing, mazzuco2021measure}. Therefore, this paper aims to bring a contribution to this discussion, delivering a new parametric model capable of decomposing mortality through mixing models in a frequentest framework. \cite{mazzuco2021measure} proposes an approach similar to the one proposed in this paper, however the authors use a Bayesian framework.

As we have seen, the proposed model fits well the mortality curve, specially above age 100, and this model does it without overparametrization, as \cite{pollard_model}. Furthermore, as it is a mixture model, the model is flexible to become the Gompertz model ($p=0$), or the Exponential model ($p=1$). When $0<p<1$, the model fits a mortality curve with inflexion point (mortality deceleration) and plateau (mortality plateau).

The use of Brazilian mortality data shed light on the performance of the model in a low quality database. We could see that the mixture-based model captures the dynamics of mortality well only when there is a drop in mortality rates, serving as an alternative to models that do not have this characteristic.

Although the present work presents a model capable of capturing the specific dynamics of the force of mortality in certain populations, it also sheds light on other problems to be solved. Since the model is based on mixtures of distributions, we are interested in deriving hypothesis tests on the estimated parameters. One of the main ones is to test if $p = 0$, i.e. whether the model can be reduced to a Gompertz model; similar interest to that studied in \cite{bohnstedt2019detecting}, when a hypothesis test for Gamma heterogeneity is derived, and important statistical properties are studied.

Finally, in the recently published paper \cite{vaupel2022pull} point out that estimating senescence mortality is of fundamental importance to understand the pace of human aging, human longevity and how far we can live. In this sense, this work brought a method capable of identifying and estimating senescent mortality, without having a great computational cost, often seen in Bayesian analysis (See \cite{barber2015rate}), or overparameterized models, as seen in \cite{pollard_model}.

\bibliographystyle{apalike2}

\bibliography{refs}

\begin{thebibliography}{}

\bibitem[Barber et~al., 2015]{barber2015rate}
Barber, S., Voss, J., \& Webster, M. (2015).
\newblock The rate of convergence for approximate bayesian computation.
\newblock {\em Electronic Journal of Statistics}, 9(1), 80--105.

\bibitem[Barbi et~al., 2018]{barbi2018plateau}
Barbi, E., Lagona, F., Marsili, M., Vaupel, J.~W., \& Wachter, K.~W. (2018).
\newblock The plateau of human mortality: Demography of longevity pioneers.
\newblock {\em Science}, 360(6396), 1459--1461.

\bibitem[Beard, 1959]{beard1959note}
Beard, R.~E. (1959).
\newblock Note on some mathematical mortality models.
\newblock In {\em Ciba Foundation Symposium-The Lifespan of Animals (Colloquia
  on Ageing)}, volume~5  (pp.\ 302--311).: Wiley Online Library.

\bibitem[Black et~al., 2017]{black2017methuselah}
Black, D.~A., Hsu, Y.-C., Sanders, S.~G., Schofield, L.~S., \& Taylor, L.~J.
  (2017).
\newblock The methuselah effect: The pernicious impact of unreported deaths on
  old-age mortality estimates.
\newblock {\em Demography}, 54(6), 2001--2024.

\bibitem[B{\"o}hnstedt \& Gampe, 2019]{bohnstedt2019detecting}
B{\"o}hnstedt, M. \& Gampe, J. (2019).
\newblock Detecting mortality deceleration: Likelihood inference and model
  selection in the gamma-gompertz model.
\newblock {\em Statistics \& Probability Letters}, 150, 68--73.

\bibitem[Brillinger et~al., 1986]{brillinger1986natural}
Brillinger, D.~R. et~al. (1986).
\newblock The natural variability of vital rates and associated statistics.
\newblock {\em Biometrics}, 42(4), 693--734.

\bibitem[Cutler et~al., 2006]{cutler2006determinants}
Cutler, D., Deaton, A., \& Lleras-Muney, A. (2006).
\newblock The determinants of mortality.
\newblock {\em Journal of economic perspectives}, 20(3), 97--120.

\bibitem[Feehan, 2018]{feehan2018separating}
Feehan, D.~M. (2018).
\newblock Separating the signal from the noise: evidence for deceleration in
  old-age death rates.
\newblock {\em Demography}, 55(6), 2025--2044.

\bibitem[Finkelstein, 2009]{finkelstein2009understanding}
Finkelstein, M. (2009).
\newblock Understanding the shape of the mixture failure rate (with engineering
  and demographic applications).
\newblock {\em Applied Stochastic Models in Business and Industry}, 25(6),
  643--663.

\bibitem[Gavrilov \& Gavrilova, 2019]{gavrilov2019late}
Gavrilov, L.~A. \& Gavrilova, N.~S. (2019).
\newblock Late-life mortality is underestimated because of data errors.
\newblock {\em PLoS biology}, 17(2), e3000148.

\bibitem[Gomes \& Turra, 2009]{gomes2009number}
Gomes, M. M.~F. \& Turra, C.~M. (2009).
\newblock The number of centenarians in brazil: indirect estimates based on
  death certificates.
\newblock {\em Demographic Research}, 20, 495--502.

\bibitem[Gompertz, 1825a]{gompertz1825xxiv}
Gompertz, B. (1825a).
\newblock Xxiv. on the nature of the function expressive of the law of human
  mortality, and on a new mode of determining the value of life contingencies.
  in a letter to francis baily, esq. frs \&c.
\newblock {\em Philosophical transactions of the Royal Society of London},
  (115), 513--583.

\bibitem[Gompertz, 1825b]{gompertz1825}
Gompertz, B. (1825b).
\newblock Xxiv. on the nature of the function expressive of the law of human
  mortality, and on a new mode of determining the value of life contingencies.
  in a letter to francis baily, esq. frs \&c.
\newblock {\em Philosophical transactions of the Royal Society of London},
  0(115), 513--583.

\bibitem[Gonzaga \& Schmertmann, 2016]{gonzaga2016estimating}
Gonzaga, M.~R. \& Schmertmann, C.~P. (2016).
\newblock Estimating age-and sex-specific mortality rates for small areas with
  topals regression: an application to brazil in 2010.
\newblock {\em Revista Brasileira de Estudos de Popula{\c{c}}{\~a}o}, 33,
  629--652.

\bibitem[Graunt, 1662]{graunt1977natural}
Graunt, J. (1662).
\newblock Natural and political observations mentioned in a following index,
  and made upon the bills of mortality.
\newblock In {\em Mathematical Demography}  (pp.\ 11--20). Springer.

\bibitem[Heligman \& Pollard, 1980]{pollard_model}
Heligman, L. \& Pollard, J.~H. (1980).
\newblock The age pattern of mortality.
\newblock {\em Journal of the Institute of Actuaries}, 107(1), 49--80.

\bibitem[Horiuchi et~al., 2013]{horiuchi2013modal}
Horiuchi, S., Ouellette, N., Cheung, S. L.~K., \& Robine, J.-M. (2013).
\newblock Modal age at death: lifespan indicator in the era of longevity
  extension.
\newblock {\em Vienna Yearbook of Population Research}, (pp.\ 37--69).

\bibitem[Horiuchi \& Wilmoth, 1997]{horiuchi1997age}
Horiuchi, S. \& Wilmoth, J.~R. (1997).
\newblock Age patterns of the life table aging rate for major causes of death
  in japan, 1951--1990.
\newblock {\em The Journals of Gerontology Series A: Biological Sciences and
  Medical Sciences}, 52(1), B67--B77.

\bibitem[Makeham, 1860]{makeham1860law}
Makeham, W.~M. (1860).
\newblock On the law of mortality and the construction of annuity tables.
\newblock {\em Journal of the Institute of Actuaries}, 8(6), 301--310.

\bibitem[Mazzuco et~al., 2021]{mazzuco2021measure}
Mazzuco, S.~S., Suhrcke, M.~M., \& Zanotto, L.~L. (2021).
\newblock How to measure premature mortality? a proposal combining
  “relative” and “absolute” approaches.
\newblock {\em Population health metrics}, 19(1), 1--14.

\bibitem[Mirjalili, 2019]{mirjalili2019genetic}
Mirjalili, S. (2019).
\newblock Genetic algorithm.
\newblock In {\em Evolutionary algorithms and neural networks}  (pp.\ 43--55).
  Springer.

\bibitem[Nepomuceno et~al., 2019]{nepomuceno2019population}
Nepomuceno, M., Turra, C., et~al. (2019).
\newblock {\em The population of centenarians in Brazil: historical estimates
  from 1900 to 2000}.
\newblock Technical report, Max Planck Institute for Demographic Research,
  Rostock, Germany.

\bibitem[Perks, 1932]{perks1932some}
Perks, W. (1932).
\newblock On some experiments in the graduation of mortality statistics.
\newblock {\em Journal of the Institute of Actuaries}, 63(1), 12--57.

\bibitem[Pinheiro \& Queiroz, 2019]{pinheiro2019regional}
Pinheiro, P.~C. \& Queiroz, B.~L. (2019).
\newblock Regional disparities in brazilian adult mortality: an analysis using
  modal age at death (m) and compression of mortality (iqr).
\newblock {\em Anais}, (pp.\ 1--20).

\bibitem[Queiroz et~al., 2020]{queiroz2020comparative}
Queiroz, B.~L., Gonzaga, M.~R., Vasconcelos, A., Lopes, B.~T., \& Abreu, D.~M.
  (2020).
\newblock Comparative analysis of completeness of death registration, adult
  mortality and life expectancy at birth in brazil at the subnational level.
\newblock {\em Population health metrics}, 18(1), 1--15.

\bibitem[Remund et~al., 2017]{remund2017analyzing}
Remund, A., Camarda, C.~G., \& Riffe, T. (2017).
\newblock Analyzing the young adult mortality hump in r with morthump.
\newblock {\em Rostock: Max Planck Institute for Demographic Research (MPIDR
  Technical Report TR-2018-003)}.

\bibitem[Remund et~al., 2018]{remund2018cause}
Remund, A., Camarda, C.~G., \& Riffe, T. (2018).
\newblock A cause-of-death decomposition of young adult excess mortality.
\newblock {\em Demography}, 55(3), 957--978.

\bibitem[Scrucca, 2013]{alg_genetic}
Scrucca, L. (2013).
\newblock {GA}: A package for genetic algorithms in {R}.
\newblock {\em Journal of Statistical Software}, 53(4), 1--37.

\bibitem[van Raalte, 2021]{van2021have}
van Raalte, A.~A. (2021).
\newblock What have we learned about mortality patterns over the past 25 years?
\newblock {\em Population Studies}, 75(sup1), 105--132.

\bibitem[Vaupel et~al., 2022]{vaupel2022pull}
Vaupel, J.~W. et~al. (2022).
\newblock {\em The Pull of the Plateau and the Sway of the Mode: Formal
  Relationships to Estimate the Pace of Senescence}.
\newblock Technical report, Center for Open Science.

\bibitem[Vaupel et~al., 2021]{vaupel2021demographic}
Vaupel, J.~W., Villavicencio, F., \& Bergeron-Boucher, M.-P. (2021).
\newblock Demographic perspectives on the rise of longevity.
\newblock {\em Proceedings of the National Academy of Sciences}, 118(9).

\bibitem[Wachter, 2018]{wachter2018hypothetical}
Wachter, K.~W. (2018).
\newblock Hypothetical errors and plateaus: A response to newman.
\newblock {\em PLoS biology}, 16(12), e3000076.

\bibitem[Wilmoth et~al., 2012]{wilmoth2012flexible}
Wilmoth, J., Zureick, S., Canudas-Romo, V., Inoue, M., \& Sawyer, C. (2012).
\newblock A flexible two-dimensional mortality model for use in indirect
  estimation.
\newblock {\em Population studies}, 66(1), 1--28.

\bibitem[Wilmoth, 2000]{wilmoth2000demography}
Wilmoth, J.~R. (2000).
\newblock Demography of longevity: past, present, and future trends.
\newblock {\em Experimental gerontology}, 35(9-10), 1111--1129.

\bibitem[Wrigley-Field, 2014]{wrigley2014mortality}
Wrigley-Field, E. (2014).
\newblock Mortality deceleration and mortality selection: three unexpected
  implications of a simple model.
\newblock {\em Demography}, 51(1), 51--71.

\end{thebibliography}

\end{document}